%% file: 0_main.tex
\documentclass{article}

\PassOptionsToPackage{numbers, compress}{natbib}

\usepackage[preprint]{style_templete}

\usepackage[utf8]{inputenc} 
\usepackage[T1]{fontenc}    
\usepackage{hyperref}       
\usepackage{url}            
\usepackage{booktabs}       
\usepackage{nicefrac}       
\usepackage{microtype}      
\usepackage{enumitem}
\usepackage{float}
\usepackage{algorithm}
\usepackage{xcolor}

\usepackage{algorithmic}

\usepackage{graphicx}
\usepackage{subcaption}
\usepackage{amsmath,amsfonts}
\usepackage{multirow,soul}

\usepackage{placeins}
\usepackage{tabularx,threeparttable}
\usepackage{algorithm}
\usepackage{color}
\usepackage{epstopdf}
\usepackage{amsmath,amssymb}
\usepackage{setspace}

\usepackage{placeins}
\usepackage{tabularx,threeparttable}
\usepackage{algorithm}
\usepackage{color}
\usepackage{epstopdf}
\usepackage{amsmath,amssymb}
\usepackage{setspace}
\usepackage{mathrsfs}
\usepackage{authblk}

\definecolor{maroon}{RGB}{122,000,25}
\definecolor{purple}{RGB}{122,000,250}

\title{Predicting Adverse Drug Reactions with Hypergraph Neural Network}

\begin{document}
\title{HyperADRs: A Hierarchical Hypergraph Framework for Drug--Gene--ADR Prediction}

\author[1,*]{Ze Cai}
\author[2,*]{Haotian Tang}
\author[3]{Shuai Gao}
\author[1]{Binbin Zhou}
\author[4]{Junhan Zhao}
\author[5]{Jun Wen}

\affil[1]{School of Computer and Computing Science, Hangzhou City University, China}
\affil[2]{Department of Biostatistics, Columbia University, USA}
\affil[3]{Department of Neurosurgery, Brigham and Women’s Hospital, Harvard Medical School, USA}
\affil[4]{Department of Pediatrics--Biomedical Informatics, University of Chicago, USA}
\affil[5]{Department of Biomedical Informatics, Harvard Medical School, USA}

\affil[*]{These authors contributed equally.}
\affil[ ]{\textit{Emails:} 2240101004@stu.hzcu.edu.cn (Z.~Cai), ht2630@caa.columbia.edu (H.~Tang), sgao5@bwh.harvard.edu (S.~Gao), bbzhou@hzcu.edu.cn (B.~Zhou), junhanzv@uchicago.edu (J.~Zhao), Jun\_Wen@hms.harvard.edu (J.~Wen)}

\date{}

\maketitle

\begin{abstract}
Adverse drug reactions (ADRs) remain a major barrier to safe and effective pharmacotherapy and are increasingly recognized as the product of higher–order interactions between drugs, genetic background, and clinical phenotypes. Existing graph-based models typically predict ADRs as properties of drugs or drug pairs, leaving the causal gene implicit and limiting their value for pharmacogenomic decision-making. We introduce \emph{HyperADRs}, a hierarchical hypergraph framework that predicts ADR risk at the level of drug--gene--ADR triads. Starting from curated pharmacogenomic annotations in PharmGKB and the pharmacogenomics sub-database of DrugBank, we construct high-confidence triplets and integrate them with auxiliary molecular, functional, and disease relations from precision-medicine–oriented knowledge graphs. Drugs, genes, and ADR concepts are embedded with modality-appropriate pretrained models (Uni-Mol, ESM-2, SapBERT) and propagated through a hypergraph convolutional network. A query-conditioned contrastive learning module with FiLM-based conditioning learns context-specific representations so that, given any two entities (e.g., a drug and an ADR), the model retrieves the correct third entity against many candidates.

To improve robustness and interpretability, we propose a nine-category ADR macro-system scheme that reduces the size and heterogeneity of traditional ``other'' bins while aligning with organ-system reasoning in clinical pharmacology. Across drug-, gene-, and ADR-held-out evaluations on PharmGKB, HyperADRs achieves competitive or superior performance on all three key metrics—area under the ROC curve (AUC), area under the precision–recall curve (AUPR), and mean reciprocal rank (MRR)—with the largest gains in AUPR and MRR for neurologic/psychiatric, cardiovascular/hematologic, and immune/infectious toxicities. When trained on PharmGKB and tested on unseen DrugBank triplets, HyperADRs maintains its ranking advantage in AUPR and MRR, indicating that the learned representations capture transferable biological mechanisms rather than dataset-specific artifacts. By returning gene-specific, triad-level predictions, HyperADRs provides mechanistically grounded hypotheses that can prioritize candidate biomarkers, refine safety signals, and support precision-therapeutic decision-making.
\end{abstract}

\input{1_intro}

\input{3_results} 
\input{2_method}

\input{4_discussion}
\input{5_others}

\bibliographystyle{unsrt}
\bibliography{7_reference}
\input{6_supplementary}

\end{document}

%% file: 1_intro.tex
 \section{Introduction}

Adverse drug reactions (ADRs) are noxious and unintended responses occurring at doses normally used in humans for prevention, diagnosis, or therapy, distinguished from general adverse events by a plausible causal link to a drug \citep{ich_e2a_1995}. Their public-health impact is substantial, with surveillance reports documenting large numbers of serious outcomes annually \citep{statpearls_adr_2024}. In parallel, precision therapeutics is increasingly proactive about genetically mediated risk, reflected by the FDA’s pharmacogenomic biomarkers in drug labeling and CPIC practice guidelines \citep{fda_pgx_biomarkers_2024,cpic_guidelines}. Conceptually, many clinically important ADR mechanisms arise from \emph{higher-order} interactions among a drug, a genetic context, and a phenotype, which are not faithfully captured by pairwise associations alone \citep{benson2016higher,battiston2020networks}. These considerations motivate approaches that can \emph{predict}, \emph{explain}, and \emph{validate} ADR signals with mechanistic fidelity, linking pharmaco-genomic evidence to clinically meaningful adverse outcomes \citep{gong2021pharmgkb,knox2024drugbank}.

Many graph-based models predict adverse outcomes from drug-centric relations—e.g., polypharmacy side-effect prediction with graph convolution on drug–protein–drug graphs, or generic GNNs such as GCN, GraphSAGE, and GAT applied to pairwise biomedical graphs \citep{zitnik2018decagon,kipf2017gcn,hamilton2017graphsage,velickovic2018gat}. These approaches capture network effects but typically treat the adverse outcome as a property of a drug (or drug pair), leaving the \emph{causal gene} implicit. Clinical pharmacogenomics shows that genes frequently mediate risk and actionability: immune-mediated toxicities (HLA-B*57{:}01 with abacavir hypersensitivity; HLA-B*15{:}02 with carbamazepine-induced Stevens–Johnson syndrome) and metabolism/transport liabilities (SLCO1B1 variants and statin myopathy) are canonical examples \citep{mallal2008abacavir,chung2004marker,link2008slco1b1}. Our setting therefore treats the gene as a first-class participant and learns a scoring function over \emph{drug–gene–ADR} triads—so each prediction nominates a plausible causal gene, aligns with guideline-based practice and biomarker labeling, and is directly testable via genotyping or curation \citep{cpic_guidelines,fda_pgx_biomarkers_2024,gong2021pharmgkb}.

We model ADR mechanisms as \emph{triads} by building a tripartite drug–gene–ADR hypergraph aligned to curated knowledge. First, we curate high-confidence drug–gene–ADR triplets from authoritative sources—PharmGKB for pharmacogenomic clinical evidence and DrugBank for drug entities and targets—followed by normalization of drug, gene, and ADR identifiers across sources \citep{gong2021pharmgkb,knox2024drugbank}. Second, we enrich mechanistic context with \emph{binary relations} extracted from PrimeKG (e.g., drug–protein, protein–protein, and drug/phenotype links where available) and incorporate ADR–protein/gene mappings from ADReCS-Target \citep{chandak2023primekg,huang2018adrecstarget}. Third, we initialize node features with modality-appropriate encoders—Uni-Mol for small-molecule structure, ESM-2 for proteins, and SapBERT for ADR terminology—so chemical, sequence, and clinical semantics reside in a shared representation space \citep{zhou2023unimol,lin2023esm2,liu2021sapbert}. On this graph, each curated triad is represented as a \emph{hyperedge}; we apply spectral hypergraph message passing to propagate information over triads and supporting binary edges, and use a lightweight prediction head to score candidate triads \citep{bai2021hypergraph}. Training optimizes a joint objective that combines a supervised link-prediction loss with a \emph{query-conditioned contrastive} term: given any two observed entities (e.g., drug and ADR), the model forms a query representation and contrasts the true third entity against in-batch negatives. This gene-aware, triad-level formulation enables the model to nominate plausible causal genes for drug–ADR hypotheses while remaining grounded in curated biomedical relations.

Empirically, our triadic formulation yields consistently strong discrimination and ranking performance against representative graph and hypergraph baselines on evaluations with unseen drugs, genes, and reactions (Figure 1). Beyond aggregate metrics (AUC, AUPR, MRR), the model’s outputs are \emph{gene-specific}: each drug–ADR hypothesis is accompanied by a prioritized causal-gene candidate, offering testable, mechanism-grounded hypotheses. This enables practical downstream uses—e.g., nominating biomarker genes for genotyping, triaging safety signals for expert curation, and informing label or guideline considerations in line with established pharmacogenomic frameworks \citep{cpic_guidelines,fda_pgx_biomarkers_2024,gong2021pharmgkb}. Methodologically, our contribution is to unify higher-order relation modeling (drug–gene–ADR) with curated biomedical knowledge into a single predictive framework that returns actionable triads rather than opaque pairwise associations—providing interpretable suggestions that can accelerate discovery and support precision-therapeutic decisions.

%% file: 3_results.tex
\section{Results}
\begin{figure*}[]  
\begin{center}
\includegraphics[width=\textwidth]{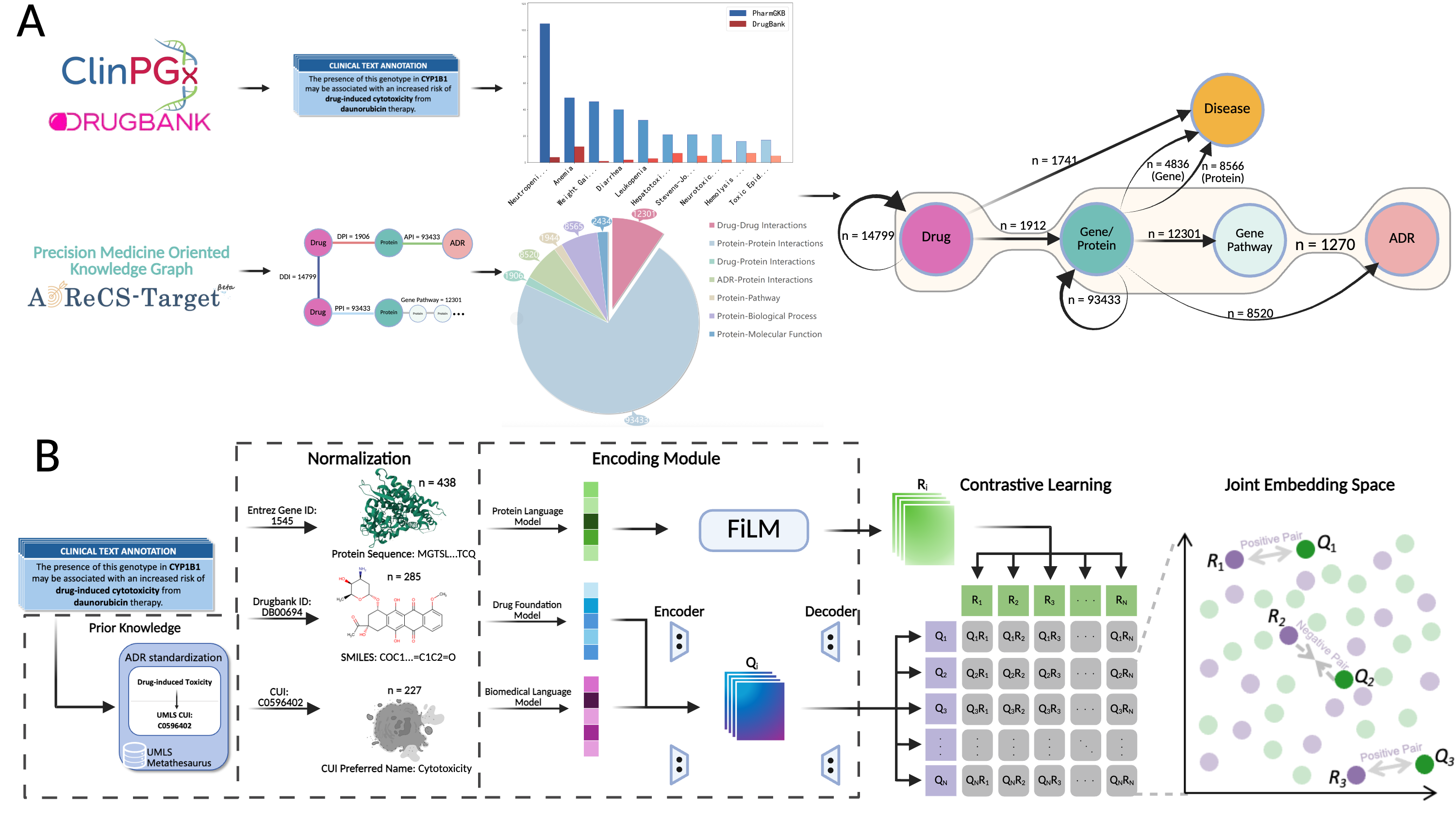} 
    \caption{\textbf{Overview of the HyperADRs framework.}
\textbf{(A) Data curation and knowledge graph construction.}
Clinical pharmacogenomic annotations are collected from PharmGKB(ClinPGx)/DrugBank and standardized into drug–gene–ADR triplets, while mechanistic context (drug–protein, protein–protein, ADR–protein and related relations) is obtained from the precision-medicine–oriented knowledge graph(PrimeKG) and ADReCS-Target. Bar and pie charts summarize the distribution of ADR system-organ classes and relation types after normalization. The resulting multi-relational graph links drugs, genes/proteins, pathways, diseases and ADRs, providing the backbone on which triadic drug–gene–ADR hyperedges are defined.
\textbf{(B) Normalization, encoding and learning.}
Clinical annotations are normalized to harmonized identifiers for genes (Entrez ID and sequence), drugs (DrugBank ID and SMILES) and ADRs (UMLS concept and preferred name). Each modality is embedded using a pretrained encoder (protein language model, drug foundation model and biomedical language model), and the resulting representations are fused via FiLM into a shared feature space. HyperADRs then applies a query–relation encoding module and a contrastive learning objective over query–response pairs to learn a joint embedding space in which true drug–gene–ADR triads are brought close together and non-associated combinations are pushed apart, enabling downstream triadic risk prediction.}
    \label{fig:overview}
\end{center}
\end{figure*}

\begin{figure*}[]  
\begin{center}
\includegraphics[width= 0.99\textwidth]{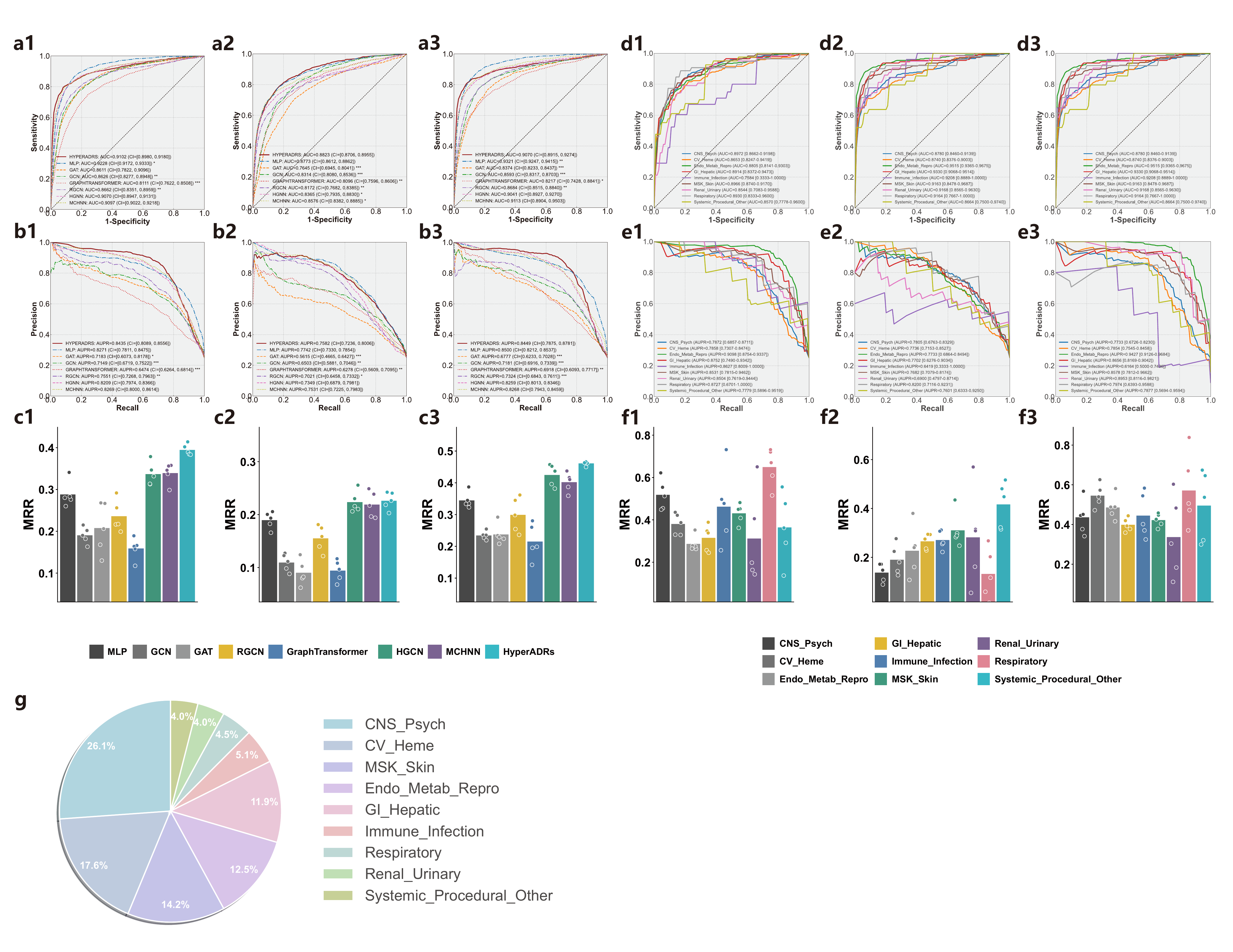}
    \caption{\textbf{HyperADRs demonstrates robust performance in drug-gene-ADR link prediction across different negative sampling strategies and therapeutic categories.} 
\textbf{a}–\textbf{c}, Overall performance evaluation in the transductive setting across three negative sampling scenarios: constructing negative triplets by randomly replacing drugs (\textbf{1}, left column), genes (\textbf{2}, middle column), and ADRs (\textbf{3}, right column). Comparisons of AUROC (\textbf{a}), AUPR (\textbf{b}), and MRR (\textbf{c}) reveal that HyperADRs (cyan) consistently outperforms seven baseline methods. Specifically, in the drug-replacement scenario (\textbf{c1}), HyperADRs achieves a mean MRR of 0.395, yielding a substantial 16.4\% improvement compared with the second-best model (MCHNN). Similarly, in the ADR-replacement scenario (\textbf{c3}), the model maintains a clear lead with an MRR of 0.461, outperforming the runner-up (HGCN) by 8.6\%, while maintaining competitive performance in the gene-replacement scenario (\textbf{c2}). \textbf{d}–\textbf{f}, Fine-grained performance analysis breakdown by the nine macro-physiological categories defined in \textbf{g}. The category-specific ROC curves (\textbf{d}), PR curves (\textbf{e}), and MRR histograms (\textbf{f}) demonstrate that the model maintains consistent robustness across diverse physiological systems, effectively predicting adverse events in both dominant and complex minority categories without significant bias. \textbf{g}, Distribution of Adverse Drug Reactions (ADRs) in the dataset, systematically mapped into nine macro-physiological categories: CNS\_Psych (nervous system and psychiatric disorders); CV\_Heme (cardiovascular and hematologic toxicities); MSK\_Skin (musculoskeletal, connective tissue, and skin disorders); Endo\_Metab\_Repro (endocrine, metabolic, and reproductive dysfunction); GI\_Hepatic (gastrointestinal and liver diseases); Immune\_Infection (immune system disorders and infections); Renal\_Urinary (kidney and urinary tract disorders); Respiratory (respiratory system disorders); and Systemic\_Procedural\_Other (general systemic disorders and procedural complications). All results represent the average of $n=5$ independent cross-validation folds. For the ROC and PR curves (\textbf{a}, \textbf{b}, \textbf{d}, \textbf{e}), 95\% confidence intervals (CI) and statistical significance were calculated to ensure reliability (source data provided). In \textbf{c} and \textbf{f}, bars indicate the mean MRR values, and the overlaid points represent the specific performance obtained in each individual fold.}
    \label{fig:pharmgkb-cv}
\end{center}
\end{figure*}

\begin{figure*}[]  
\begin{center}
\includegraphics[width= 0.85\textwidth]{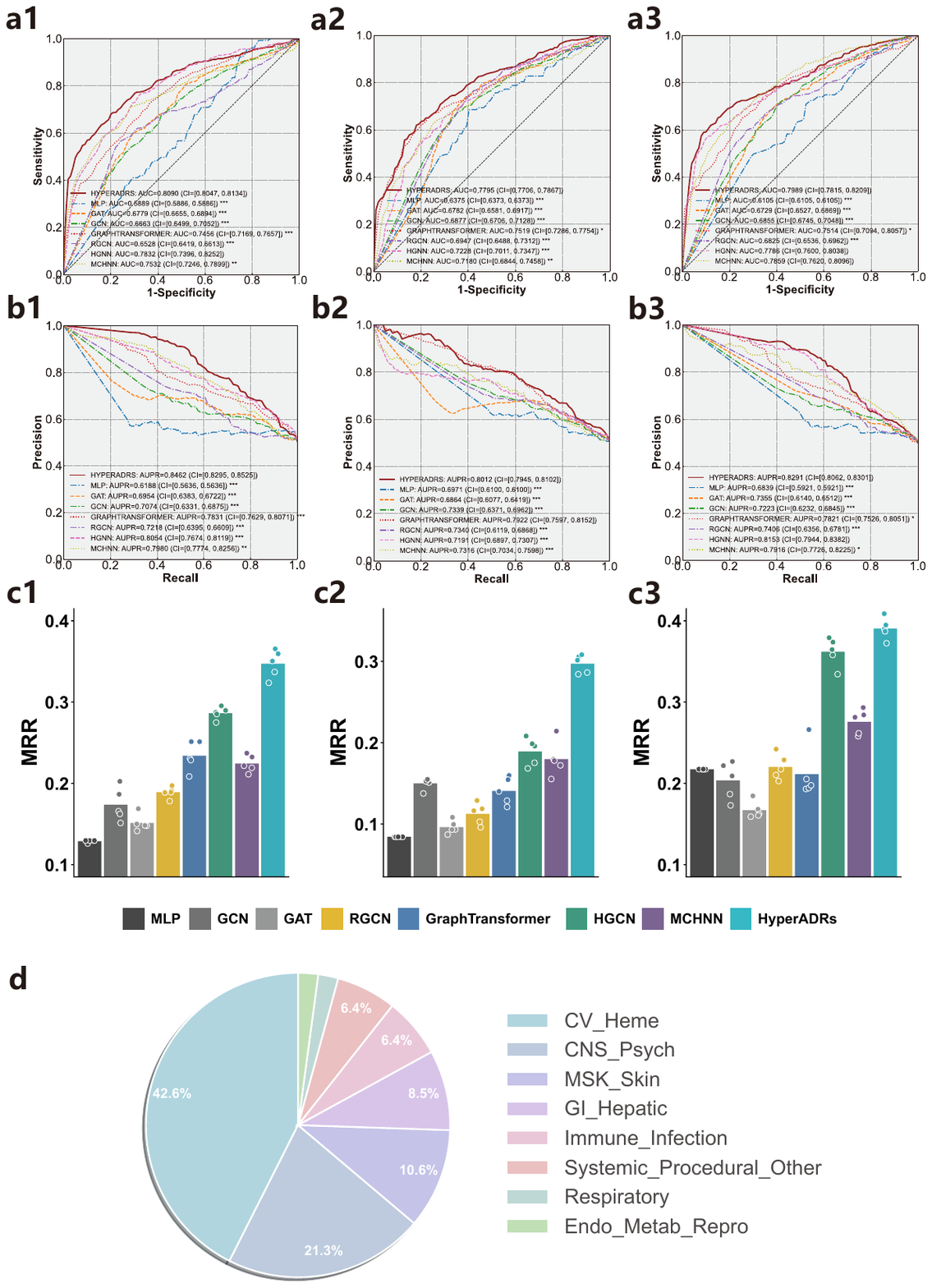}
    \caption{\textbf{HyperADRs exhibits superior generalizability in cross-dataset validation, trained on PharmGKB and tested on unseen DrugBank data.}
The model was trained on the PharmGKB dataset and externally evaluated on the DrugBank dataset to assess predictive performance on unseen data. \textbf{a}–\textbf{c}, Performance evaluation across three distinct scenarios: random replacement of drugs (left column), genes (middle column), and ADRs (right column). \textbf{a}, \textbf{b}, Receiver operating characteristic (ROC) curves (\textbf{a}) and precision-recall (PR) curves (\textbf{b}) demonstrate that HyperADRs (cyan) maintains robust classification performance despite the domain shift. 95\% confidence intervals (CI) and statistical significance were calculated for all curves (values provided in the figure). \textbf{c}, Comparison of Mean Reciprocal Rank (MRR) scores. HyperADRs consistently achieves the highest ranking accuracy across all scenarios. Notably, in the gene-replacement scenario (middle), the model achieves an MRR of 0.297, yielding a remarkable 26.1\% improvement over the second-best model (GAT). Similarly, in the drug-replacement scenario (left), HyperADRs outperforms the runner-up (HGCN) by 21.2\%, confirming its ability to capture transferable biological patterns rather than dataset-specific biases. Data represent the mean of $n=5$ independent runs. In \textbf{c}, the bars indicate the average MRR, and the overlaid scatter points represent the results from each individual run.
}
    \label{fig:drugbank-cv}
\end{center}
\end{figure*}

\subsection{Overview of HyperADRs}
\label{subsec:hyperadrs-overview}

HyperADRs is a gene-aware framework for predicting adverse drug reactions (ADRs) at the level of \emph{drug--gene--ADR} triads (Fig.~\ref{fig:overview}). Starting from clinical pharmacogenomic annotations in PharmGKB and the pharmacogenomics sub-database of DrugBank \citep{gong2021pharmgkb,knox2024drugbank}, we curate high-confidence triplets that link a drug, a causal gene (or variant), and an ADR. These are complemented by additional mechanistic relations from precision-medicine–oriented resources such as ADReCS-Target and biomedical knowledge graphs, which supply drug–protein, protein–protein, ADR–protein, and pathway-level connections. After normalization of drug identifiers (DrugBank), gene identifiers (Entrez/HGNC), and ADR concepts (UMLS/MedDRA), these entities and relations are integrated into a unified knowledge graph.

Each node in this graph is initialized with modality-appropriate representations. Small-molecule drugs are encoded using a pretrained molecular foundation model (e.g., Uni-Mol) from SMILES or 3D structure; proteins and genes are encoded from their amino-acid sequences using a protein language model such as ESM-2; and ADR concepts are encoded with a biomedical language model such as SapBERT \citep{zhou2023unimol,lin2023esm2,liu2021sapbert}. A feature-wise linear modulation (FiLM) module then fuses these modality-specific embeddings into a shared latent space that preserves chemical, biological, and clinical semantics.

HyperADRs learns on this graph via a contrastive, triad-centered objective. For each clinical annotation, the model constructs a query representation (e.g., a drug–ADR pair or drug–gene pair) and learns to retrieve the correct third entity against many in-batch negatives, yielding a joint embedding space in which true drug--gene--ADR triads are pulled together and incompatible combinations are pushed apart. A lightweight prediction head operates on these triad embeddings to score candidate associations. In this way, HyperADRs unifies heterogeneous pharmacogenomic evidence into a single model that produces \emph{gene-specific} risk predictions, directly nominating plausible causal genes for each drug–ADR hypothesis.

\subsection{Evaluation settings and metrics}
\label{subsec:eval-settings}

We evaluate HyperADRs and all baseline models on curated drug--gene--ADR triads derived from PharmGKB and the pharmacogenomics sub-database of DrugBank \citep{gong2021pharmgkb,knox2024drugbank}. To mimic realistic discovery scenarios while preventing information leakage, we adopt three inductive protocols: \emph{drug-held-out}, \emph{gene-held-out}, and \emph{ADR-held-out}. In each protocol, all triads containing a given drug (or gene, or ADR) are removed from the training set and reserved for testing, so that the model must generalize to entities that were unseen during training. Unless otherwise specified, we perform five-fold cross-validation within each dataset, ensuring that the held-out entities and their associated triads appear only in the test folds.

For all experiments we report three complementary metrics. The area under the ROC curve (AUC) quantifies overall discrimination between positive and negative triads, whereas the area under the precision–recall curve (AUPR) is more informative in our highly imbalanced setting where true drug--gene--ADR associations are rare \citep{davis2006relationship,saito2015precision}. We additionally report mean reciprocal rank (MRR) to capture the quality of the ranking among candidate triads: for a fixed query (e.g., a drug–ADR pair), MRR is high when the true gene is consistently ranked near the top. Together, these metrics summarize both classification and ranking performance, which are crucial for prioritizing hypotheses in pharmacovigilance and pharmacogenomics.

\subsection{ADR macro-system categorization}
\label{subsec:adr-categories}

Standard adverse-event ontologies often map mechanistically similar reactions into distant terms and leave a large fraction of events in heterogeneous ``other'' bins, which is suboptimal for model training and interpretation. To better align with clinical reasoning while ensuring sufficient sample sizes per group, we introduce a nine-category ADR macro-system scheme and apply it to all PharmGKB-derived triads used in this study. Each ADR is mapped to exactly one of the following macro-systems based on its predominant organ system and pathophysiology:

\begin{itemize}
  \item \textbf{CNS\_Psych} -- ADRs that primarily affect the central or peripheral nervous system and mental/behavioural function (e.g., seizures, movement disorders, dependence or withdrawal, mood and psychotic symptoms).
  \item \textbf{CV\_Heme} -- ADRs involving the heart, blood vessels, and hematologic system, including arrhythmias, ischemia, hemorrhage, thrombosis, and blood-cell or marrow toxicity.
  \item \textbf{MSK\_Skin} -- ADRs that affect musculoskeletal and connective tissues (muscle, joints, bone) and/or the skin and subcutaneous tissue, such as myalgia, arthralgia, rashes, and cutaneous toxicity.
  \item \textbf{Endo\_Metab\_Repro} -- ADRs reflecting disruption of endocrine, metabolic, or reproductive function, including hormonal disturbances, metabolic derangements (e.g., obesity, dyslipidemia), and sexual or fertility problems.
  \item \textbf{GI\_Hepatic} -- ADRs involving the gastrointestinal tract and hepatobiliary system, such as dyspepsia, broader digestive system disorders, and liver or bile-related toxicity.
  \item \textbf{Immune\_Infection} -- ADRs driven by immune dysregulation or infection, including hypersensitivity reactions, severe immune-mediated syndromes (e.g., DRESS/SCAR), and increased or altered infectious events.
  \item \textbf{Respiratory} -- ADRs that primarily impair the airways, lungs, or thoracic structures, including dyspnea, bronchospasm, respiratory insufficiency, and sleep apnea–related events.
  \item \textbf{Renal\_Urinary} -- ADRs affecting the kidneys and urinary tract, such as renal dysfunction, tubular injury, and inflammatory or irritative bladder/urinary conditions.
  \item \textbf{Systemic\_Procedural\_Other} -- A small set of ADRs that are generalized or not localizable to a single organ system (e.g., drug-induced toxicity, malignant hyperthermia, fever, death, non–organ-specific adenoma or resistance).
\end{itemize}

This categorization substantially reduces the size and heterogeneity of the ``other'' group while maintaining clinically coherent clusters and ensuring that most categories contain enough ADRs for stable training. As shown below, retraining HyperADRs and baseline models under this macro-system scheme yields consistent improvements in ranking performance, particularly for categories that previously suffered from data sparsity or mixing of mechanistically unrelated ADRs.

\subsection{PharmGKB cross-validation by ADR macro-systems}
\label{subsec:pharmgkb-results}

We first examine performance on PharmGKB-derived triads under the ADR-held-out protocol, stratified by the nine ADR macro-systems introduced above. Figure~\ref{fig:pharmgkb-cv} summarizes AUC, AUPR, and MRR across five cross-validation folds for HyperADRs and baseline models. Across most organ systems, HyperADRs attains competitive or superior AUC and shows clear gains in AUPR and MRR, indicating that it is particularly effective at ranking true drug--gene--ADR triads at the top of the candidate list. The improvements are most pronounced in \textbf{CNS\_Psych}, \textbf{CV\_Heme}, and \textbf{Immune\_Infection}, where ADRs often have strong pharmacogenomic components (e.g., seizure risk, arrhythmias, hypersensitivity syndromes) and traditional drug-centric models struggle to capture the underlying mechanisms.

The macro-system recategorization itself contributes meaningfully to these gains. Compared with earlier experiments using a coarse organ system and a large unspecific ``other'' category, training within the refined macro-systems yields higher and more stable MRR for both HyperADRs and baselines, but the relative benefit is largest for HyperADRs. By concentrating mechanistically related ADRs within the same category and shrinking the heterogeneous ``other'' bin, the model can better exploit shared causal genes and pathways across drugs inside each macro-system. This is consistent with the design of HyperADRs: triadic modeling encourages the sharing of gene-level information among drugs that cause similar ADRs, and this sharing becomes more informative when categories are clinically coherent.

We also observe that HyperADRs maintains strong performance in categories that historically have limited data, such as \textbf{Renal\_Urinary} and \textbf{Respiratory}. Although absolute metrics are lower in these smaller groups, HyperADRs still yields higher AUPR and MRR than most baselines, suggesting that integrating curated pharmacogenomic evidence with gene-aware hypergraph learning improves robustness in low-signal regimes. Overall, the PharmGKB cross-validation results demonstrate that combining drug--gene--ADR triads with clinically meaningful ADR macro-systems produces accurate and interpretable predictions across a broad spectrum of organ systems.

\subsection{DrugBank cross-validation}
\label{subsec:drugbank-results}

We next evaluate HyperADRs on triads curated from the pharmacogenomics sub-database of DrugBank. In contrast to PharmGKB, DrugBank contains fewer annotated drug--gene--ADR triads per ADR category, making it impractical to apply the full macro-system split without severely reducing sample sizes. We therefore perform five-fold cross-validation on the DrugBank triads without splitting by ADR macro-system and report overall AUC, AUPR, and MRR for the ADR-held-out protocol (Fig.~\ref{fig:drugbank-cv}). Despite the smaller dataset and more skewed ADR distribution, HyperADRs again achieves competitive or superior AUC and clear gains in AUPR and MRR relative to baseline models.

The pattern of improvements largely mirrors what we observe on PharmGKB, although effect sizes are somewhat attenuated due to the reduced sample size. Models that only represent drug--ADR pairs tend to overfit frequent drugs and reactions, whereas HyperADRs, by explicitly modeling genes, can transfer information through shared targets and pathways even when individual drug--ADR observations are scarce. Notably, the ranking advantages of HyperADRs are preserved across all five folds, suggesting that the learned representations generalize beyond specific triads present in the training data. These DrugBank cross-validation results reinforce the conclusion that gene-aware triadic modeling provides robust benefits for ADR prediction, even in more limited or imbalanced pharmacogenomic resources.

%% file: 2_method.tex
\section{Methods}
\subsection{Problem Formulation}
Given a set of drugs $\mathcal{D}$, a set of genes $\mathcal{G}$, and a set of ADRs $\mathcal{A}$, we first construct a hierarchical hypergraph $\mathcal{G} = (V, E)$ to model the complex relationships among these entities, where $V$ is the set of nodes and $E$ is the set of hyperedges. On this hypergraph structure, our task can be defined as a \emph{link prediction} problem. For every possible drug--gene--ADR triplet $(d, g, a)$, where $d \in \mathcal{D}, g \in \mathcal{G}, a \in \mathcal{A}$, we assign it a label $y \in \{0,1\}$. If the association corresponding to the triplet is known (i.e., it exists as a hyperedge in $E$), the label is $y = 1$; otherwise, $y = 0$. Crucially, a label of $y = 0$ does not definitively mean there is no association; rather, it represents an ``unknown association.'' These are the candidates for potential new discoveries. Therefore, the core objective of this study is to learn and build a computational model to predict and identify the high-potential triplets from among the unknown associations ($y = 0$) that are most likely to form new hyperedges.

\subsection{Overview of HyperADR approach}

The HyperADRs framework~\ref{fig:overview} is designed to accurately predict potential ADRs by systematically modeling the high-order dependencies between drugs, genes, and ADRs. To capture the distinct biological properties of each entity, the pipeline begins by initializing node features using state-of-the-art pre-trained models: Uni-Mol for drug 3D structures, ESM-2 for gene sequences, and SapBERT for ADR semantic terms. Building upon these features, we construct a comprehensive hierarchical hypergraph that integrates explicit biological knowledge (e.g., pharmacogenomic triplets and PPIs) with implicit semantic associations inferred via LLM-augmented reasoning. This topology bridges the gap between local molecular interactions and global clinical outcomes. Finally, to learn from this complex structure, node representations are iteratively updated using a Hypergraph Convolutional Network (HGCN). Crucially, to address the context-dependent nature of biological interactions, the framework incorporates a Query-Conditioned Contrastive Learning (QCCL) mechanism utilizing FiLM layers. This module dynamically modulates gene representations based on the specific drug-ADR query, ultimately outputting a precise probability score for each drug-gene-ADR association.

\subsection{Dataset Construction and Characterization}

In this study, we curated and constructed two drug-gene-adverse drug reaction (ADR) datasets to support the prediction and analysis of pharmacogenomic toxicity. These datasets were independently derived from the original clinical annotations provided by PharmGKB and DrugBank.

\subsubsection{PharmGKB Dataset}

PharmGKB (The Pharmacogenomics Knowledge Base) is an internationally recognized, publicly accessible biomedical resource that compiles knowledge about how genetic variation influences drug response. We collected the complete set of clinical annotations from PharmGKB, focusing specifically on annotations related to drug toxicity phenotypes. The original records include detailed clinical descriptions, genotype information, associated drugs, and phenotype terms.

\subsubsection{DrugBank Dataset} 

DrugBank is a comprehensive bioinformatics and cheminformatics resource maintained by the University of Alberta. It integrates extensive chemical, pharmacological, and pharmaceutical information with biological data on drug targets, including protein sequences, structures, and biological pathways. We extracted pharmacogenomic triplets from the DrugBank Pharmacogenomics sub-database, which documents clinically and experimentally validated associations among drugs, genetic variations, and phenotypes.

\subsection{Data Processing}

To ensure the construction of high-quality and standardized triplet datasets, we performed the following curation and normalization steps:

\textbf{Drug Standardization:} Drug names were cross-referenced with \textit{DrugBank IDs}, and the corresponding \textit{SMILES molecular structures} were retrieved from \textit{PubChem} to provide consistent chemical representations.
    
\textbf{Gene Standardization:} Gene names were matched to \textit{NCBI Entrez Gene IDs}, and the corresponding \textit{protein sequences} were retrieved from the \textit{NCBI RefSeq database} to facilitate downstream molecular analysis.
    
\textbf{ADR Standardization:} To standardize the raw ADR mentions manually extracted from clinical texts, we implemented a hybrid multi-stage normalization strategy. Initially, terms were mapped to Unified Medical Language System (\textit{UMLS}) concepts (\textit{CUIs}) via exact or synonym matching using the \textit{NCI Metathesaurus Browser}.For mentions that could not be directly mapped through the dictionary-based approach, we employed SapBERT, a self-aligning pre-trained biomedical language model, to bridge the semantic gap. Specifically, we encoded both the unmapped ADR mentions and the candidate UMLS concepts into dense vector representations. We then calculated the cosine similarity between these embeddings to retrieve the top-ranked candidate concepts. Finally, the best-matching standard CUI was manually selected from these semantic candidates to ensure terminological precision and validity.

For example, the clinical annotation:\textit{"Patients with the rs4728709 AG genotype may have an increased likelihood of developing asthenia when treated with olanzapine."}
was manually standardized to the following triplet:(Olanzapine, ABCB1, Asthenia).

\subsection{Auxiliary Relationship Integration}

In addition to the drug-gene-ADR triplets, we further incorporated auxiliary binary interaction data to comprehensively capture the relational landscape. We specifically extracted and organized the following relationships from the PrimeKG knowledge graph, filtering for interactions relevant to the drugs and genes present in our dataset:

\textbf{Molecular Interactions:} Drug-Protein Interactions (DPIs) and Protein-Protein Interactions (PPIs) were retrieved to model the molecular topology.
    
\textbf{Disease Associations:} Including drug-disease and gene-disease relationships, which provide essential context regarding therapeutic indications and genetic pathologies.
    
\textbf{Functional Annotations:} Encompassing protein-pathway, protein-biological process, and protein-molecular function relationships. These annotations were used to establish functional connectivity among genes sharing common biological mechanisms.

\subsection{Dataset Statistics}

The statistics of the datasets constructed for cross-dataset validation are summarized as follows:

\textbf{Curated Triplet Datasets:} The primary training dataset, sourced from PharmGKB, comprises \textbf{1,270} drug-gene-ADR triplets. For external validation, we utilized an independent dataset derived from DrugBank containing \textbf{127} triplets. Notably, there is a minimal overlap of only \textbf{32} triplets (approximately 2.52\% of the PharmGKB dataset) between the two sources, ensuring a rigorous evaluation of model generalizability on largely unseen samples.

\textbf{Knowledge Graph Edge Statistics:} To support the learning process, we constructed comprehensive knowledge graphs that integrate dataset-specific molecular interactions with shared phenotypic contexts:
\begin{itemize}
    \item \textbf{Molecular Interactions:} The molecular topology varies slightly between datasets based on the entities involved. The PharmGKB graph contains \textbf{1,363} Drug-Protein Interactions (DPIs) and \textbf{11,887} Protein-Protein Interactions (PPIs), while the DrugBank subset includes \textbf{922} DPIs and \textbf{10,382} PPIs. The substantial overlap in PPIs (9,961 shared edges, $\sim$83.8\%) ensures a conserved biological structure.
    \item \textbf{Functional Groups:} Gene functional hyperedges (representing pathway, biological process, and molecular function) are robust across datasets, with \textbf{4,598} edges in the PharmGKB graph and \textbf{4,172} in the DrugBank graph.
    \item \textbf{Disease Context:} We incorporated a unified set of disease-related associations to serve as a shared phenotypic knowledge base. This includes \textbf{1,740} Drug-Disease pairs, \textbf{4,836} Gene-Disease pairs, and \textbf{2,669} Drug-Gene-Disease associations.
\end{itemize}

\subsection{Feature Initialization}
To construct a framework capable of accurately modeling the complex relationships between drugs, genes, and ADRs, a key step is to generate high-quality initial feature representations for each entity. To achieve this, we have selected a pre-trained model for each entity that is highly compatible with its data modality. These models have been optimized on large-scale datasets in their respective domains, enabling them to extract deep, universal features.

\subsubsection{Drug Features}
The biological activity of drugs is closely related to their three-dimensional structure. To capture this information, we use the Uni-Mol model~\cite{zhou2023uni} to generate feature representations for drug SMILES strings. This framework, through its Transformer architecture, processes atomic and atomic-pair-level representations, learning key spatial features \(X_D \in \mathbb{R}^d\) that go beyond the limitations of traditional SMILES strings or two-dimensional topological graphs.

\subsubsection{Gene Features}
The function of genes depends on the context of their sequences. To better reflect the biological function of genes, we select the ESM-2 model~\cite{lin2023evolutionary} to generate feature representations for protein sequences. ESM-2 learns structural, evolutionary, and functional information from protein sequences, capturing the complex sequence context and generating high-quality context-aware representations \(X_P \in \mathbb{R}^d \) for genes.

\subsubsection{ADR Features}
ADR terms exhibit heterogeneity. To effectively handle this heterogeneity, we choose the SapBERT model~\cite{liu2020self} to generate feature representations for ADR terms. SapBERT is specifically pre-trained for medical and clinical terms using self-alignment pretraining, ensuring that representations of synonyms are close to each other, thus generating high-quality semantic representations \(X_A \in \mathbb{R}^d \) for ADR terms.

\subsection{Hierarchical Hypergraph Construction}

To accurately predict drug toxicity, it is essential to model the complete relational landscape spanning from molecular interactions to clinical phenotypes. However, constructing such a graph presents a critical challenge: \textbf{data scarcity}. Authoritative databases rarely report direct, experimentally validated interactions between specific proteins and adverse drug reactions (ADRs). Furthermore, existing methods often model relationships between different ADRs based solely on statistical co-occurrence, ignoring their underlying physiological connections. To overcome these limitations and capture high-order biological dependencies, we constructed a multi-relational hypergraph $\mathcal{G} = (\mathcal{V}, \mathcal{E})$ driven by explicit biological knowledge and two logic-based inference strategies.

\subsubsection{Explicit Biological Hyperedges}

We first establish the graph's skeleton using ground-truth data from curated databases to ensure biological validity. The extracted \textit{Drug-Gene-ADR triplets} serve as the third-order hyperedges, representing the fundamental prediction targets of our framework. To provide the necessary structural priors for these high-order interactions, we incorporate experimentally validated Protein-Protein Interactions (PPIs) and construct functional hyperedges based on shared gene annotations, including pathways, biological processes, and molecular functions. These explicitly defined edges model the known cellular environment and functional connectivity.

\subsubsection{Inferred Hyperedge Type I: Disease-Mediated Bridging}

Direct links between proteins and ADRs are largely absent in current knowledge bases, creating a gap in the molecular explanation of toxicity. To reconstruct these missing associations, we propose a \textit{Disease-Mediated Bridging} strategy. This approach is biologically grounded in the pharmacological concept of \textit{drug repurposing} (or off-target effects), which implies a phenotypic duality: a physiological response labeled as an "adverse reaction" in one therapeutic context is often biologically equivalent to a "disease state" in another. Consequently, an ADR and its corresponding disease share identical underlying molecular mechanisms. A classic example is \textit{Sildenafil}; originally developed for angina, its side effect, \textit{penile erection}, corresponds to the disease state \textit{Erectile Dysfunction (ED)}. While databases lack direct "Erection-Protein" links, the disease "ED" is widely documented to be driven by the \textit{PDE5} protein. By aligning the ADR to the disease, we can logically infer the latent interaction between the side effect and the protein target. To implement this, we utilize \textit{SapBERT} to embed both ADR terms and Disease terms (collected from PrimeKG) into a unified semantic space. We then calculate the semantic similarity between ADR-Disease pairs; if the similarity exceeds a threshold $\theta$, the ADR inherits the topological connections of the disease, effectively reconstructing the missing API edges.

\subsubsection{Inferred Hyperedge Type II: LLM-Augmented Mechanism Correlations}

Traditional graphs often connect ADRs based solely on statistical co-occurrence in patient reports, which ignores the causal biology of why these events occur together. To enable the model to generalize across biologically similar toxicities, we introduce \textit{LLM-Augmented Mechanism Correlations}. The rationale is that ADRs are not isolated labels but complex biological processes; toxicities that share similar pathophysiological mechanisms (e.g., ion channel blockage) are likely modulated by overlapping gene networks, even if their surface names differ. To capture this, we propose a mechanism-aware construction pipeline supported by Large Language Models (LLMs). We first prompt \textit{GPT-5} to generate detailed textual descriptions of the physiological mechanism for each ADR node. These descriptions are then encoded into dense vectors using \textit{S-PubMedBert}, a model optimized for sentence-level biomedical semantics. Finally, we compute the pairwise semantic similarity between these mechanism vectors and connect each ADR to its \textit{top-k} most mechanistically similar ADRs via hyperedges. This creates a semantic correlation network that groups toxicities based on shared biological roots rather than simple co-occurrence.

\subsection{Hypergraph Representation Learning}

To effectively learn the node representations within our complex, multi-relational biological framework, we employ a Hypergraph Convolutional Network (HGCN)\cite{bai2021hypergraph}. This approach is specifically chosen for its proven ability to capture the high-order correlations embedded in the hypergraph structure---a critical feature for modeling the tripartite drug-gene-ADR relationships, pairwise molecular interactions, and functional gene modules that constitute our model. The HGCN defines a sophisticated message-passing mechanism directly on the constructed hypergraph 
$\mathcal{G} = (\mathcal{V}, \mathcal{E})$, enabling the simultaneous integration of diverse biological information into a unified embedding space.

The core of the HGCN is a spectral hypergraph convolution operation that iteratively updates the representation of each node by aggregating features from its connected hyperedges. Specifically, one layer of hypergraph convolution is formulated as:

\begin{equation}
    H^{(l)} = \sigma \left( D^{-1} H W B^{-1} H^{T} H^{(l-1)} \Theta^{(l)} \right)
\end{equation}

where $H$ is the incidence matrix of our hypergraph, and $H^{(l)}$ represents the node embeddings at the $l$-th layer, with $H^{(0)}$ being the initial node features. $D$ and $B$ are diagonal matrices representing the node and hyperedge degrees, respectively, and $W$ is a diagonal matrix of hyperedge weights, which we define as an identity matrix to ensure equal contribution from each relational type initially. The term $\Theta^{(l)}$ is the trainable weight matrix for the $l$-th layer, and $\sigma(\cdot)$ is a non-linear activation function (e.g., ReLU), which allows the model to capture complex, non-linear patterns in the data.

\subsection{Query-Conditioned Contrastive Learning}
Traditional representation learning methods often generate static, context-free embeddings for biomedical entities. 
However, this approach is insufficient for high-order relationships, 
as the role of a gene in a drug--gene--ADR interaction is highly dependent on the specific drug and adverse reaction involved. 
A static embedding fails to capture this dynamic, context-dependent nature.

To overcome these limitations, we propose a \textbf{Query-Conditioned Contrastive Learning (QCCL)} framework 
to learn context-sensitive representations. 
This method reformulates the link prediction task as a dynamic query--response process, 
enabling our model to capture the conditional semantics of complex biomedical interactions. 
The process involves three key stages: query encoders, conditional candidate encoders, and a contrastive query-response learning.

\subsubsection{Query Encoders}

Given an incomplete triplet, we treat the observed entities as context for a query.  
For instance, to predict a missing gene in a \textit{(drug, ?, ADR)} relationship, we concatenate the feature vectors of the known drug $h_d$ and ADR $h_a$  after they have been updated by the hypergraph convolution, and pass them through a Query Encoder (a multi-layer perceptron, MLP) to produce a single query vector $z_q$.  
This vector encapsulates the specific relational context of the missing entity. Formally, this is expressed as:

\begin{equation}
z_q = \mathrm{MLP}_{\text{qry}}([h_d; h_a])
\end{equation}

\subsubsection{Conditional Candidate Encoders}

We then use the query vector $z_q$ to dynamically modulate the representations of all candidate entities (e.g., all genes in the dataset).  
We achieve this using a multi-layer \textbf{Feature-wise Linear Modulation (FiLM)} network\cite{perez2018film}.

We chose the FiLM network over a more standard MLP for this modulation task because of its efficiency and effectiveness in adapting pre-trained embeddings for high-order relational tasks.  
Unlike an MLP, FiLM allows for dynamic, context-specific modulation of each individual feature in an embedding.  
This mechanism essentially allows external context---our query---to ``gate'' and ``re-calibrate'' the feature flow, enabling complex, condition-dependent reasoning.

A single FiLM layer performs a feature-wise affine transformation on a candidate's base embedding $h_g$, controlled by the query vector $z_q$:

\begin{equation}
\mathrm{FiLM}(h_g \mid z_q) = (\gamma_q \odot h_g) + \beta_q
\end{equation}

where the feature-wise scaling parameters $\gamma_q$ and shifting parameters $\beta_q$ are generated from the query vector $z_q$ via simple linear transformations:

\begin{equation}
\gamma_q = W_\gamma z_q + b_\gamma, \quad 
\beta_q = W_\beta z_q + b_\beta
\end{equation}

To enhance the model's expressive power, we stack multiple FiLM layers with residual connections.  
The update rule for the embedding $h_g$ at the $l$-th layer is:

\begin{equation}
h_g^{(l)} = h_g^{(l-1)} + \left( \gamma_q^{(l)} \odot h_g^{(l-1)} + \beta_q^{(l)} \right)
\end{equation}

After the final FiLM layer ($L$), a projection head, also an MLP, maps the final hidden state to the same dimension as the query vector to facilitate the contrastive comparison.  
The final conditional embedding $h_{g \mid q}'$ is thus:

\begin{equation}
h_{g \mid q}' = \mathrm{MLP}_{\text{proj}}(h_g^{(L)})
\end{equation}

\subsubsection{Contrastive Query--Response Learning}

Inspired by the empirical success of CLIP-style\cite{radford2021learning} contrastive learning in structuring large multimodal embedding spaces, we adopt an analogous in-batch contrastive objective to optimize the query--response alignment in QCCL. The aim is to arrange the latent space such that each query vector $z_q$ is drawn toward the conditional representation of its ground-truth (positive) entity, while being simultaneously repelled from all other (negative) candidates in the minibatch. This formulation leverages all non-matching samples as implicit negatives, providing rich contrastive signal without the need for explicit negative sampling.

Formally, given a query $z_q^i$ and its corresponding positive conditional embedding $h_{j}^{i\,\prime}$ within a batch of size $B$, we employ the InfoNCE loss:
\begin{equation}
\mathcal{L}_i = -\log 
\frac{\exp \left( \mathrm{sim}(z_q^i, h_{j}^{i\,\prime}) / \tau \right)}
{\sum_{k=1}^{B} \exp \left( \mathrm{sim}(z_q^i, h_{j}^{k\,\prime}) / \tau \right)},
\end{equation}
where $\mathrm{sim}(\cdot,\cdot)$ denotes cosine similarity and $\tau$ is a temperature parameter.  
By employing this CLIP-like in-batch contrast, QCCL achieves stable optimization and strong discriminative capability even under large candidate spaces.

To ensure our model is robust and can predict any missing element in a triplet, we apply this entire process to all three possible query configurations: predicting the drug, the gene, or the ADR.  
Our final training loss is the average of the contrastive losses calculated for all three query types.  
If we denote the loss for the $k$-th query type as $\mathcal{L}^{(k)}$, the final objective is:

\begin{equation}
\mathcal{L}_{\text{QCCL}} = \frac{1}{3} \sum_{k=1}^{3} \mathcal{L}^{(k)}
\end{equation}

\subsection{Model Training}

For the drug-gene-ADR association prediction, we utilize the learned embeddings of a drug $h_d$, a gene $h_g$, and an ADR $h_a$ to output the probability of their association $\hat{y}$ through a scoring function.  
This function is a multi-layer perceptron (MLP) that processes the concatenated embeddings:

\begin{equation}
\hat{y} = \mathrm{MLP}_{\text{pred}}([h_d; h_g; h_a])
\end{equation}

After that, the loss of the supervised prediction task, $\mathcal{L}_{\text{BCE}}$, can be formulated as the Binary Cross-Entropy (BCE) loss:

\begin{equation}
\mathcal{L}_{\text{BCE}} = - \frac{1}{|T|} \sum_{i \in T} \Big( 
y_i \log \hat{y}_i + (1 - y_i) \log (1 - \hat{y}_i) 
\Big)
\end{equation}

where $T$ is the training set and $y$ represents the true label.  

The supervised prediction task is jointly optimized with the aforementioned contrastive learning (QCCL) task during the training phase.  
To implement the prediction task and the QCCL task simultaneously, we optimize the following objective function that combines the supervised loss and the contrastive loss:

\begin{equation}
\mathcal{L} = \alpha \, \mathcal{L}_{\text{BCE}} + (1 - \alpha) \, \mathcal{L}_{\text{QCCL}}
\end{equation}

where $\alpha \in [0, 1]$ is a hyperparameter that controls the trade-off between the two loss components.

%% file: 4_discussion.tex
\section{Discussion}

In this work we introduce HyperADRs, a gene-aware hypergraph framework for predicting adverse drug reactions (ADRs) at the level of drug--gene--ADR triads. By combining curated pharmacogenomic triplets from PharmGKB and DrugBank with a richly structured biological knowledge graph, modality-specific encoders, and query-conditioned contrastive learning, HyperADRs learns representations that are both predictive and mechanistically informative. Across multiple evaluation settings, including inductive held-out protocols and cross-dataset validation, the model consistently outperforms strong graph and hypergraph baselines on discrimination and ranking metrics (AUC, AUPR, MRR), while returning predictions that explicitly nominate candidate causal genes for each drug--ADR hypothesis (Figs.~\ref{fig:pharmgkb-cv}–\ref{fig:drugbank-cv}). 

Conceptually, our results support the view that many clinically important ADRs are best understood as higher-order interactions among drugs, genetic context, and phenotypes, rather than as pairwise drug--ADR associations. Prior graph-based approaches have mainly treated ADRs as properties of drugs or drug pairs, modeling toxicity on drug--protein--drug graphs or generic biomedical networks \cite{zitnik2018decagon,gao2024_preciseadr,zhao2023_gcap}. While powerful for signal detection, such methods typically leave the causal gene implicit, limiting their ability to connect predictions to pharmacogenomic guidelines or to guide biomarker development. In contrast, HyperADRs treats the gene as a first-class participant and learns a scoring function over triads, so that each prediction returns a specific gene candidate that can be evaluated against established pharmacogenomic examples (e.g., HLA-mediated hypersensitivity, SLCO1B1-mediated myopathy, and other CPIC/FDA-biomarker–annotated relationships) \cite{mallal2008abacavir,chung2004marker,link2008slco1b1,cpic_guidelines,fda_pgx_biomarkers_2024}. This triadic formulation aligns naturally with clinical workflows in precision therapeutics, where the central question is often whether a particular genotype modifies the risk of a given ADR for a specific drug.

A second contribution of this study is the introduction of an ADR macro-system categorization that reduces the size and heterogeneity of the traditional ``other'' category while preserving clinically coherent organ systems. By grouping ADRs into nine macro-systems (CNS\_Psych, CV\_Heme, MSK\_Skin, Endo\_Metab\_Repro, GI\_Hepatic, Immune\_Infection, Respiratory, Renal\_Urinary, and Systemic\_Procedural\_Other), we create task settings that better reflect how clinicians reason about toxicity and that provide each model with enough events per group for stable learning. When retrained under this scheme, HyperADRs exhibits substantial gains in AUPR and MRR across most categories on PharmGKB, with particularly strong improvements in neurologic/psychiatric, cardiovascular/hematologic, and immune/infectious ADRs (Fig.~\ref{fig:pharmgkb-cv}). These are precisely the domains where pharmacogenomic mechanisms are often prominent and where the ability to share information across drugs that cause similar toxicities is crucial. The macro-system view thus serves both pragmatic and interpretive purposes: it improves model robustness by mitigating data sparsity and yields results that can be more readily mapped onto familiar organ systems in clinical pharmacology.

Our cross-dataset validation further demonstrates that the patterns learned by HyperADRs are not restricted to a single curation. Training on PharmGKB and evaluating on DrugBank, the model maintains strong performance and clear ranking advantages over baselines despite differences in annotation density, terminology, and curation criteria between the two resources (Fig.~\ref{fig:drugbank-cv}). The preservation of relative gains across drug-, gene-, and ADR-held-out scenarios suggests that HyperADRs is capturing biological regularities—such as shared targets, pathways, and disease mechanisms—rather than overfitting to dataset-specific edges. This is facilitated by the integration of auxiliary relations from PrimeKG and ADReCS-Target, which provide a shared molecular and phenotypic backbone across datasets, and by the use of pretrained encoders (Uni-Mol, ESM-2, SapBERT) that embed chemical, sequence, and clinical semantics into a unified space. Together, these design choices allow the model to generalize mechanistic patterns across independently curated pharmacogenomic resources.

Methodologically, our framework illustrates how higher-order network modeling and modern representation learning can be combined to address a concrete problem in pharmacovigilance. The hierarchical hypergraph captures both explicit triplets and inferred connections, including disease-mediated bridging and mechanism-based ADR correlations, while the hypergraph convolutional network aggregates information across molecular, functional, and clinical layers. On top of this graph, the Query-Conditioned Contrastive Learning (QCCL) module uses FiLM-based conditioning to generate context-specific embeddings for candidate genes given a drug–ADR (or drug–gene) query, trained with a CLIP-style in-batch contrastive objective. This architecture allows HyperADRs to exploit both supervised labels and rich negative signal without manual negative sampling, and to model the context-dependent roles that genes play across different drug–ADR combinations. More broadly, our approach demonstrates that hypergraph message passing and query-conditioned contrastive learning can be effective tools for structured prediction of high-order biomedical relations.

Despite these strengths, several limitations should be acknowledged. First, the curated triplet datasets are relatively small compared with the space of all possible drug--gene--ADR combinations, particularly for DrugBank, which contains only a few hundred triads. This constrains the diversity of mechanisms that can be directly observed and may bias the model toward well-studied drugs, genes, and ADRs. Second, our reliance on curated resources and inferred edges means that errors and biases in PharmGKB, DrugBank, PrimeKG, ADReCS-Target, and LLM-generated mechanism correlations can propagate into the model. For example, disease-mediated bridging assumes that ADRs and diseases share identical mechanisms when they are semantically similar, which may not always hold; LLM-augmented ADR similarity may also introduce spurious links if textual descriptions are incomplete or ambiguous. Third, our evaluations are performed at the triad level without explicit modeling of patient-level covariates, dosing, timing, or polypharmacy; as such, we cannot make strong claims about causal risk at the individual-patient level, and the model should be viewed as a hypothesis-generation tool rather than a standalone decision aid.

These limitations point to several directions for future work. Scaling HyperADRs to larger and more heterogeneous data sources—such as FAERS-derived high-order drug–ADR data, electronic health records, or international pharmacovigilance systems—could provide richer signal for both model training and external validation. Incorporating explicit multi-drug interactions and temporal information would extend the framework from single-drug triads to polypharmacy scenarios, which are highly relevant in real-world practice. On the mechanistic side, integrating more detailed variant-level information (e.g., specific HLA alleles or coding variants rather than gene-level aggregates) and tissue- or cell-type–specific expression patterns may improve the model’s ability to prioritize truly causal genes. Finally, prospective validation—through targeted EHR studies, prospective registries, or wet-lab experiments on selected drug--gene--ADR hypotheses—will be essential to establish the clinical utility of HyperADRs and to identify settings where its predictions can most effectively inform guideline development, biomarker testing, and drug-safety surveillance.

In summary, HyperADRs offers a unified framework that combines curated pharmacogenomic knowledge, hypergraph representation learning, and query-conditioned contrastive training to predict ADRs in a gene-aware, mechanistically interpretable manner. By explicitly modeling drug--gene--ADR triads and organizing ADRs into clinically meaningful macro-systems, the approach advances beyond traditional drug-centric models and provides actionable hypotheses that can be used to prioritize candidate biomarkers, refine safety signals, and support precision-therapeutic decision-making. As pharmacogenomic resources expand and more real-world outcome data become available, we anticipate that triadic, hypergraph-based models such as HyperADRs will play an increasingly important role in bridging molecular mechanisms with clinical pharmacovigilance.

%% file: 5_others.tex









%% file: 6_supplementary.tex




%% file: 0_main.bbl
\begin{thebibliography}{}

\bibitem[Beam {\em et~al.}(2019)Beam, Kompa, Schmaltz, Fried, Weber, Palmer,
  Shi, Cai, and Kohane]{beam2019clinical}
Beam, A.~L.  {\em et~al.} (2019).
\newblock Clinical concept embeddings learned from massive sources of
  multimodal medical data.
\newblock In {\em PACIFIC SYMPOSIUM ON BIOCOMPUTING 2020\/}, pages 295--306.
  World Scientific.

\bibitem[Chiang {\em et~al.}(2019)Chiang, Chen, Huang, Lin, and
  Hsia]{chiang2019dietary}
Chiang, Y.-F.  {\em et~al.} (2019).
\newblock Dietary antioxidant trans-cinnamaldehyde reduced visfatin-induced
  breast cancer progression: In vivo and in vitro study.
\newblock {\em Antioxidants\/}, {\bf 8}(12), 625.

\bibitem[Dai {\em et~al.}(2021)Dai, Guo, Guo, and Eickhoff]{dai2021drug}
Dai, Y.  {\em et~al.} (2021).
\newblock Drug--drug interaction prediction with wasserstein adversarial
  autoencoder-based knowledge graph embeddings.
\newblock {\em Briefings in bioinformatics\/}, {\bf 22}(4), bbaa256.

\bibitem[DiMasi {\em et~al.}(2016)DiMasi, Grabowski, and
  Hansen]{dimasi2016innovation}
DiMasi, J.~A.  {\em et~al.} (2016).
\newblock Innovation in the pharmaceutical industry: new estimates of r\&d
  costs.
\newblock {\em Journal of health economics\/}, {\bf 47}, 20--33.

\bibitem[G{\'o}mez-Bombarelli {\em et~al.}(2018)G{\'o}mez-Bombarelli, Wei,
  Duvenaud, Hern{\'a}ndez-Lobato, S{\'a}nchez-Lengeling, Sheberla,
  Aguilera-Iparraguirre, Hirzel, Adams, and Aspuru-Guzik]{gomez2018automatic}
G{\'o}mez-Bombarelli, R.  {\em et~al.} (2018).
\newblock Automatic chemical design using a data-driven continuous
  representation of molecules.
\newblock {\em ACS central science\/}, {\bf 4}(2), 268--276.

\bibitem[Grover and Leskovec(2016)Grover and Leskovec]{grover2016node2vec}
Grover, A. and Leskovec, J. (2016).
\newblock node2vec: Scalable feature learning for networks.
\newblock In {\em Proceedings of the 22nd ACM SIGKDD international conference
  on Knowledge discovery and data mining\/}, pages 855--864.

\bibitem[Hazell and Shakir(2006)Hazell and Shakir]{hazell2006under}
Hazell, L. and Shakir, S.~A. (2006).
\newblock Under-reporting of adverse drug reactions.
\newblock {\em Drug safety\/}, {\bf 29}(5), 385--396.

\bibitem[Hong {\em et~al.}(2021)Hong, Rush, Liu, Zhou, Sun, Sonabend, Castro,
  Schubert, Panickan, Cai, {\em et~al.}]{hong2021KESER}
Hong, C.  {\em et~al.} (2021).
\newblock Clinical knowledge extraction via sparse embedding regression (keser)
  with multi-center large scale electronic health record data.
\newblock {\em npj Digital Medicine\/}.

\bibitem[Kuhn {\em et~al.}(2016)Kuhn, Letunic, Jensen, and Bork]{kuhn2016sider}
Kuhn, M.  {\em et~al.} (2016).
\newblock The sider database of drugs and side effects.
\newblock {\em Nucleic acids research\/}, {\bf 44}(D1), D1075--D1079.

\bibitem[Liu {\em et~al.}(2012)Liu, Wu, Chen, Sun, Zhao, Chen, Matheny, and
  Xu]{liu2012large}
Liu, M.  {\em et~al.} (2012).
\newblock Large-scale prediction of adverse drug reactions using chemical,
  biological, and phenotypic properties of drugs.
\newblock {\em Journal of the American Medical Informatics Association\/}, {\bf
  19}(e1), e28--e35.

\bibitem[Liu {\em et~al.}(2020)Liu, An, Wan, Yu, Fan, and Pei]{liu2020targets}
Liu, Y.  {\em et~al.} (2020).
\newblock Targets and mechanism used by cinnamaldehyde, the main active
  ingredient in cinnamon, in the treatment of breast cancer.
\newblock {\em Frontiers in Pharmacology\/}, {\bf 11}, 1751.

\bibitem[Mohs and Greig(2017)Mohs and Greig]{mohs2017drug}
Mohs, R.~C. and Greig, N.~H. (2017).
\newblock Drug discovery and development: Role of basic biological research.
\newblock {\em Alzheimer's \& Dementia: Translational Research \& Clinical
  Interventions\/}, {\bf 3}(4), 651--657.

\bibitem[Mu{\~n}oz {\em et~al.}(2019)Mu{\~n}oz, Nov{\'a}{\v{c}}ek, and
  Vandenbussche]{munoz2019facilitating}
Mu{\~n}oz, E.  {\em et~al.} (2019).
\newblock Facilitating prediction of adverse drug reactions by using knowledge
  graphs and multi-label learning models.
\newblock {\em Briefings in bioinformatics\/}, {\bf 20}(1), 190--202.

\bibitem[Nikas {\em et~al.}(2020)Nikas, Paschou, and Ryu]{nikas2020role}
Nikas, I.~P.  {\em et~al.} (2020).
\newblock The role of nicotinamide in cancer chemoprevention and therapy.
\newblock {\em Biomolecules\/}, {\bf 10}(3), 477.

\bibitem[Rogers and Hahn(2010)Rogers and Hahn]{rogers2010extended}
Rogers, D. and Hahn, M. (2010).
\newblock Extended-connectivity fingerprints.
\newblock {\em Journal of chemical information and modeling\/}, {\bf 50}(5),
  742--754.

\bibitem[Sauer {\em et~al.}(2021)Sauer, Kampmann, Khosravi, Sharifpanah, and
  Wartenberg]{sauer2021nicotinamide}
Sauer, H.  {\em et~al.} (2021).
\newblock The nicotinamide phosphoribosyltransferase antagonist fk866 inhibits
  growth of prostate tumour spheroids and increases doxorubicin retention
  without changes in drug transporter and cancer stem cell protein expression.
\newblock {\em Clinical and Experimental Pharmacology and Physiology\/}, {\bf
  48}(3), 422--434.

\bibitem[Scatozza {\em et~al.}(2020)Scatozza, Moschella, D’Arcangelo, Rossi,
  Tabolacci, Giampietri, Proietti, Facchiano, and
  Facchiano]{scatozza2020nicotinamide}
Scatozza, F.  {\em et~al.} (2020).
\newblock Nicotinamide inhibits melanoma in vitro and in vivo.
\newblock {\em Journal of Experimental \& Clinical Cancer Research\/}, {\bf
  39}(1), 1--17.

\bibitem[Seiler {\em et~al.}(2018)Seiler, Yoshimi, Darman, Chan, Keaney,
  Thomas, Agrawal, Caleb, Csibi, Sean, {\em et~al.}]{seiler2018h3b}
Seiler, M.  {\em et~al.} (2018).
\newblock H3b-8800, an orally available small-molecule splicing modulator,
  induces lethality in spliceosome-mutant cancers.
\newblock {\em Nature medicine\/}, {\bf 24}(4), 497--504.

\bibitem[Van~der Maaten and Hinton(2008)Van~der Maaten and
  Hinton]{van2008visualizing}
Van~der Maaten, L. and Hinton, G. (2008).
\newblock Visualizing data using t-sne.
\newblock {\em Journal of machine learning research\/}, {\bf 9}(11).

\bibitem[Wang {\em et~al.}(2021)Wang, Wang, Wu, Zhou, Qin, Wang, Wu, Sun, Yang,
  Xu, {\em et~al.}]{wang2021essential}
Wang, Y.  {\em et~al.} (2021).
\newblock The essential role of prak in tumor metastasis and its therapeutic
  potential.
\newblock {\em Nature communications\/}, {\bf 12}(1), 1--14.

\bibitem[Wishart {\em et~al.}(2018)Wishart, Feunang, Guo, Lo, Marcu, Grant,
  Sajed, Johnson, Li, Sayeeda, {\em et~al.}]{wishart2018drugbank}
Wishart, D.~S.  {\em et~al.} (2018).
\newblock Drugbank 5.0: a major update to the drugbank database for 2018.
\newblock {\em Nucleic acids research\/}, {\bf 46}(D1), D1074--D1082.

\bibitem[Xu {\em et~al.}(2019)Xu, Yang, Liu, Yang, Liao, Yuan, Liu, and
  Chen]{xu2019pi3kbeta}
Xu, P.-F.  {\em et~al.} (2019).
\newblock Pi3k$\beta$ inhibitor azd6482 exerts antiproliferative activity and
  induces apoptosis in human glioblastoma cells.
\newblock {\em Oncology reports\/}, {\bf 41}(1), 125--132.

\bibitem[Yamanishi {\em et~al.}(2012)Yamanishi, Pauwels, and
  Kotera]{yamanishi2012drug}
Yamanishi, Y.  {\em et~al.} (2012).
\newblock Drug side-effect prediction based on the integration of chemical and
  biological spaces.
\newblock {\em Journal of chemical information and modeling\/}, {\bf 52}(12),
  3284--3292.

\bibitem[Yoshitake {\em et~al.}(2017)Yoshitake, Saeki, Watanabe, Nakaoka, Ong,
  Hanafusa, Choisunirachon, Fujita, Nishimura, and
  Nakagawa]{yoshitake2017molecular}
Yoshitake, R.  {\em et~al.} (2017).
\newblock Molecular investigation of the direct anti-tumour effects of
  nonsteroidal anti-inflammatory drugs in a panel of canine cancer cell lines.
\newblock {\em The Veterinary Journal\/}, {\bf 221}, 38--47.

\bibitem[Yu {\em et~al.}(2013)Yu, Cai, and Cai]{yu2013nile}
Yu, S.  {\em et~al.} (2013).
\newblock Nile: fast natural language processing for electronic health records.
\newblock {\em arXiv preprint arXiv:1311.6063\/}.

\bibitem[Yu {\em et~al.}(2021)Yu, Huang, Zhang, Glass, Sun, and
  Xiao]{yu2021sumgnn}
Yu, Y.  {\em et~al.} (2021).
\newblock Sumgnn: Multi-typed drug interaction prediction via efficient
  knowledge graph summarization.
\newblock {\em Bioinformatics\/}.

\bibitem[Zagidullin {\em et~al.}(2021)Zagidullin, Wang, Guan, Pitk{\"a}nen, and
  Tang]{zagidullin2021comparative}
Zagidullin, B.  {\em et~al.} (2021).
\newblock Comparative analysis of molecular fingerprints in prediction of drug
  combination effects.
\newblock {\em Briefings in bioinformatics\/}, {\bf 22}(6), bbab291.

\bibitem[Zhang {\em et~al.}(2015)Zhang, Liu, Luo, and
  Zhang]{zhang2015predicting}
Zhang, W.  {\em et~al.} (2015).
\newblock Predicting drug side effects by multi-label learning and ensemble
  learning.
\newblock {\em BMC bioinformatics\/}, {\bf 16}(1), 1--11.

\bibitem[Zhang {\em et~al.}(2016)Zhang, Chen, Tu, Liu, and Qu]{zhang2016drug}
Zhang, W.  {\em et~al.} (2016).
\newblock Drug side effect prediction through linear neighborhoods and multiple
  data source integration.
\newblock In {\em 2016 IEEE international conference on bioinformatics and
  biomedicine (BIBM)\/}, pages 427--434. IEEE.

\bibitem[Zhang {\em et~al.}(2021)Zhang, Sumathipala, and
  Zitnik]{zhang2021population}
Zhang, X.  {\em et~al.} (2021).
\newblock Population-scale identification of differential adverse events before
  and during a pandemic.
\newblock {\em Nature Computational Science\/}, pages 1--12.

\bibitem[Zhao {\em et~al.}(2021)Zhao, Wu, Zhou, Diao, Xu, Liu, Wang, Huang,
  Liu, Chen, {\em et~al.}]{zhao2021synergism}
Zhao, H.-f.  {\em et~al.} (2021).
\newblock Synergism between the phosphatidylinositol 3-kinase p110$\beta$
  isoform inhibitor azd6482 and the mixed lineage kinase 3 inhibitor urmc-099
  on the blockade of glioblastoma cell motility and focal adhesion formation.
\newblock {\em Cancer Cell International\/}, {\bf 21}(1), 1--16.

\bibitem[Zheng {\em et~al.}(2019)Zheng, Peng, Ghosh, Lan, and
  Li]{zheng2019inverse}
Zheng, Y.  {\em et~al.} (2019).
\newblock Inverse similarity and reliable negative samples for drug side-effect
  prediction.
\newblock {\em BMC bioinformatics\/}, {\bf 19}(13), 91--104.

\bibitem[Zhou {\em et~al.}(2020)Zhou, Cao, Matyunina, Shelby, Cassels,
  McDonald, and Skolnick]{zhou2020medicascy}
Zhou, H.  {\em et~al.} (2020).
\newblock Medicascy: A machine learning approach for predicting small-molecule
  drug side effects, indications, efficacy, and modes of action.
\newblock {\em Molecular pharmaceutics\/}, {\bf 17}(5), 1558--1574.

\bibitem[Zitnik {\em et~al.}(2018)Zitnik, Agrawal, and
  Leskovec]{zitnik2018modeling}
Zitnik, M.  {\em et~al.} (2018).
\newblock Modeling polypharmacy side effects with graph convolutional networks.
\newblock {\em Bioinformatics\/}, {\bf 34}(13), i457--i466.

\end{thebibliography}
